# Effect of Substrate Thickness on Responsivity of Free-Membrane Bolometric Detectors

Mehdi Hosseini, Alireza Kokabi, Ali Moftakharzadeh, Mohammad A. Vesaghi, Mehdi Fardmanesh,
*Senior Member, IEEE*

*Abstract*—**The effect of substrate thickness and its Debye temperature on the bolometric response of the freestanding-membrane type of superconductor transition-edge detectors is investigated. The investigation is based on the calculation of the variation of the specific heat per unit volume and the heat conductivity using the finite-size scaling method for different Debye temperatures and micron size thickness of substrate. We also calculated the heat capacity and time constant for various materials with Debye temperatures in the range of 600-1400K. The calculated results are compared to previously reported response values obtained theoretically and experimentally for the thermal-based infrared detectors. The IR response versus substrate thickness of these types of devices for different modulation frequencies is also presented. This inspected response at low thicknesses of substrate shows anomalistic behavior with respect to the previously reported results of response analyses. Here, we also obtained the optimized substrate thickness and Debye temperatures for maximum responsivity of these types of devices.**

*Index Terms*—**Finite size effect, Superconductor bolometer, Responsivity**

## I. INTRODUCTION

SINCE the application of superconductor materials in the bolometric IR detectors, many works have been focused to improve the performance of these types of devices. Three basic types of superconductor bolometric detectors have been introduced and studied: on a solid substrate, the bolometers on thermal filament and the membrane version. The reported measurements indicate that the free-membrane type exhibits more sensitivity among these three basic types of bolometric detectors [1]. The steady state voltage optical responsivity of a bolometer is calculated previously [2].

Experimental and theoretical analysis [1, 3] show that using the thinner substrates in the free-membrane bolometers would decrease the heat capacitance and the thermal conductance which improve the sensitivity. According to the previous reports, decreasing the substrate thickness, would increase the

Manuscript received December 9, 2010.

A. Moftakharzadeh, A. Kokabi are with the School of Electrical Engineering, Sharif University of Technology, P.O. Box 11365-9363, Tehran, Iran.

A. Bozbey is with the Department of Electrical and Electronics Engineering, TOBB Economicsand Technology University, Ankara, Turkey.

M. A. Vesaghi and M. Hosseini are with the Department of Physics, Sharif University of Technology, P.O. Box 11155-9161, Tehran, Iran.

M. Fardmanesh was with the Electrical Engineering Department, Bilkent University, Ankara, Turkey. He is now with the School of Electrical Engineering, Sharif University of Technology, P.O. Box 11365-9363, Tehran, Iran. (Corresponding author: M. Fardmanesh phone: 98-21-66165920; fax: 98-21-66023261; e-mail: fardmanesh@sharif.edu).

heat conductivity in the micrometer range of thickness [4,5], and would decrease the heat conductivity in the range of hundreds of nanometers [6] through two different mechanisms. The dependency of the specific heat and the thermal conductivity to the material dimension are described by the theory of finite-size effect [7]. In this method, the total energy of the substance is considered as a sum of its surface and bulk energies. For large dimensions, the surface energy is very small compared to the bulk energy and so its effect in the total energy is negligible. However, for very thin substrates, the surface effect is significant and the term of the surface energy becomes comparable to the bulk energy. Thus for small size devices the surface energy in the calculation of the total energy cannot be neglected. The specific heat is the variation rate of total energy with respect to the temperature, and the thermal conductivity is linearly related to specific heat [8-10]. Thus, for thin substrates, the surface effects should be included in the calculation of the thermal parameters. This results in the dependency of thermal parameters on the substrate thickness, which is confirmed by the measurements [11]. In this work, the dependencies of the specific heat per unit volume and the thermal conductivity on the substrate thickness are calculated. Also, the total heat capacitance and thermal conductance of the device versus the substrate thickness are obtained.

## II. THEORETICAL MODEL

Schematic geometry description of device model is drawn in Fig. 1.

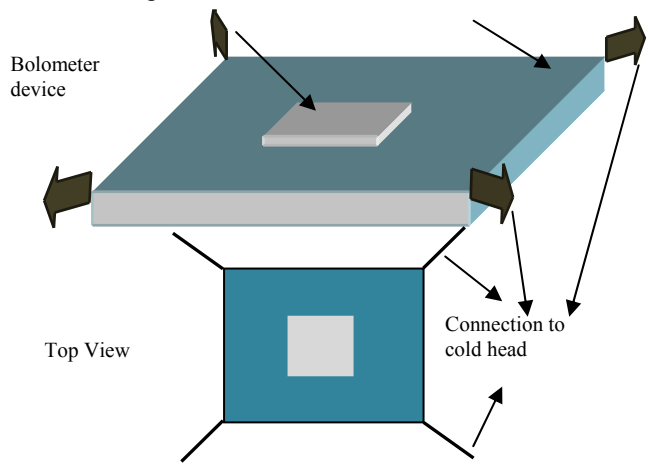

Fig. 1 Schematic picture of geometry description of device

The thermal conductance, *G* and heat capacitance, *C* of the

substrate are

$$G_s = \frac{k_s S_s}{d_s}, \; C_s = c_{Vs} S_s d_s \qquad (1)$$

where $k_s$ is the thermal conductivity, $c_{Vs}$ is the specific heat per unit volume and $S_s$ is the area of the substrate[2]. The equation 1 is valid when the substrate thickness and the lateral distance between the device area and the cold head is less than the thermal diffusion length of the substrate, $L_f$. This assumption is reliable for the membrane type bolometers. Thus, total thermal conductance of bolometer is $G_t^{-1} = G_s^{-1} + G(0)^{-1}$ in which $G(0)$ is the steady-state thermal conductance due to lateral thermal diffusion. Taking into the account the fact that $G_s \gg G(0)$, the total thermal conductance of the device is determined by $G(0)$. This $G(0)$ is also taken to be constant and determined by the thermal boundary resistance between the substrate and the temperature reservoir because the thermal diffusion length is longer than the lateral length of the substrate. By these assumptions, responsivity is calculated in this work.

The free energy of a material with finite size is found to be the sum of the bulk energy and an excess energy [12-17]. In the first order approximation, this excess energy can be attributed to the surface energy of the surrounding faces of the material. In the case of thin film, two of the surrounding faces are considerable.

Based on the above discussion, the total energy can be written as [12]

$$F(T,d) = V f_v(T) + A f_{s,1}(T) + A f_{s,2}(T) + \delta(T,d) \qquad (2)$$

where $F$ is the total free energy, $f_v$ is the bulk free energy per unit volume and $f_{s1}$ and $f_{s2}$ are the surface free energies per unit area at the two plates, $\delta$ in the higher order contribution of size effect and $d$ is the material thickness. The calculations are performed on a thin layer crystal substrate with a square surface in which $f_{s1} = f_{s2} = f_s$. In this method, considering the energies per unit volume and per unit area, the total free energy per unit volume is obtained as

$$f_{tot} = f_v + \frac{2}{d} f_s + ..., \qquad (3)$$

where $f_{tot}$ is the total energy per unit volume.

The two-dimensional internal energy per unit area, $u_{2D}$ is derived by integration of the density of states for each polarization, $D(\omega)$ over $\omega$ divided by the area of the substrate. By neglecting the contributions of phonon modes higher than the Debye frequency, $\omega_D$, the result turns out to be [3][1]

$$u_{2D} = \frac{\hbar}{\pi v^2} \int_0^{\omega_D} \frac{\omega^2 d\omega}{\exp(\hbar\omega/k_B T) - 1}, \qquad (4)$$

where the two-dimensional specific heat per unit area is found by differentiation of the $u_{2D}$ with respect to the temperature.

To evaluate the three-dimensional internal energy per unit volume, $u_{3D}$, the same strategy approached above leads to [3]

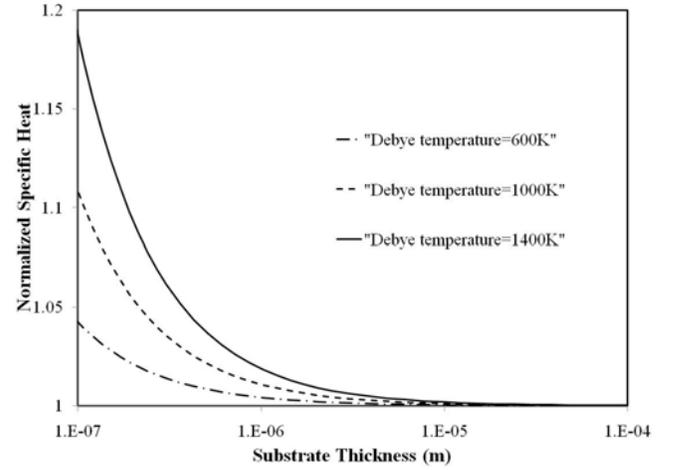

Fig. 2. Normalized specific heat with respect to thickness due to surface effects for different Debye temperatures.

$$u_{3D} = \frac{3\hbar}{2\pi^2 v^3} \int_0^{\omega_D} \frac{\omega^3 d\omega}{\exp(\hbar\omega/k_B T) - 1}. \qquad (5)$$

Now the specific heat per unit area, $C_{2D}$ and specific heat per unit volume, $C_{3D}$ can be written as:

$$c_{2D} = \left( \frac{\partial u_{2D}}{\partial T} \right)_S, \qquad (6)$$

$$c_{3D} = \left( \frac{\partial u_{3D}}{\partial T} \right)_V. \qquad (7)$$

Applying the finite-size effect to the internal energy and differentiating with respect to temperature, the total specific heat per unit volume could be obtained as

$$c_V = c_{3D} + \frac{2}{d} c_{2D}. \qquad (8)$$

### III. RESULTS AND ANALYSES

Here we present the results of our calculation for thin substrate with square active area. In Fig. 2, the normalized specific heat per unit volume versus the thickness of the substrate for different Debye temperatures is plotted. This figure shows that at large thicknesses, the specific heat per unit volume of the substrate is approximately constant, which is consistent with the previous explanation. In addition, as can be observed in the figure, the variation of the specific heat for higher Debye temperatures is larger than small Debye temperatures. Debye temperature of Silicon and Sapphire are about 650K and 1000K respectively. For large thicknesses, very small surface effect and so negligible dependency on the thickness are expected. However, for thicknesses at the range of tens of nanometers, the surface energy becomes comparable to the bulk energy and leads to increasing of the normalized heat capacitance (Fig. 2). In other word, decreasing one dimension, which results in decreasing the total volume of the substrate, makes the surface energy more significant with respect to the bulk energy. The theoretical results by Palasantzas for thin substrates [18] are consistent with our calculations. The normalized specific heat per unit volume for different thicknesses versus the Debye temperatures of the substrate is presented in Fig. 3. From this figure, it is concluded that the specific heat decreases more rapidly for higher thicknesses. Two dash lines in the figure are inverse square and inverse cubic curve that are the envelope of the specific heat. For near zero thicknesses the specific heat curve approaches to the inverse square envelope and for large thicknesses and high temperatures, the specific heat curve approaches to the inverse cubic envelope.

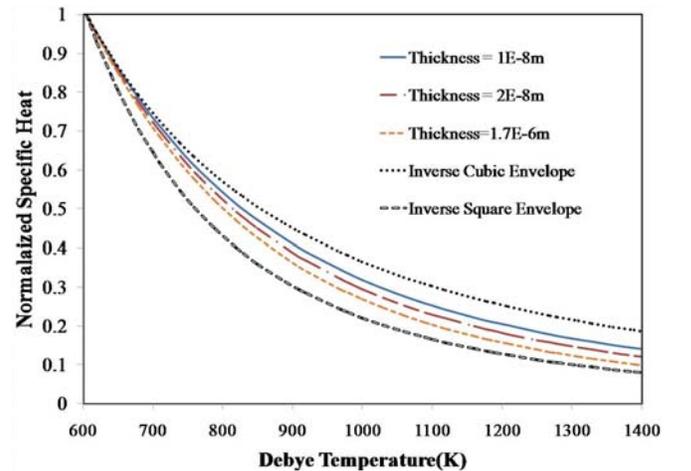

Fig. 3. Normalized specific heat with respect to Debye temperature for different thicknesses. Dash lines are the inverse square and inverse cubic envelopes.

---

[1] Note that Equation (4) and (5) were mistakenly written in Ref. [3].

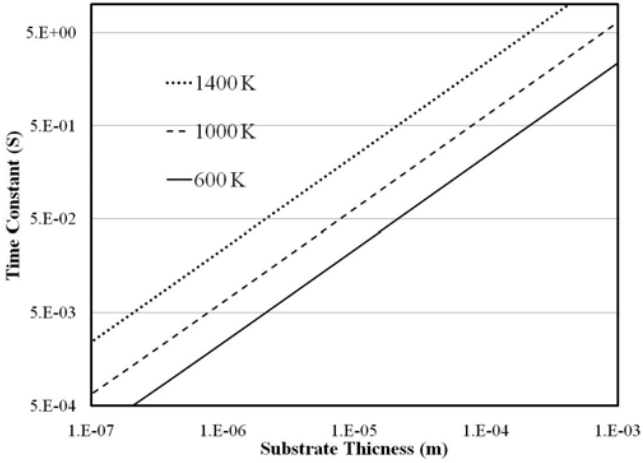

Fig. 4. Dependency of the time constant to the substrate thickness for different substrate Debye Temperatures.

It should be noted that all of the calculations are performed at the YBCO transition temperature, 90K. Considering the temperatures much lower than the Debye temperature, the volume specific heat is proportional to $T^3$ and surface specific heat is proportional to $T^2$ (due to equations 4 and 5). By decreasing temperature, volume specific heat decreases more rapidly than the surface specific heat. Therefore, for lower temperatures the surface effect becomes more significant.

Now, we repeat the calculation for a typical low temperature device with SiN substrate. We take the working temperature, substrate thickness and Debye temperature to be 6K, 120nm and 1200K, respectively. The total heat capacitance is found to be 3.4 times larger than the bulk heat capacitance. If the temperature were taken 2K, then the total heat capacitance would be 10 times larger than the bulk one.

The temperature dependence of the bulk heat capacitance are found to be in the form of $T^3$[10,19], while the one for surface have $T^2$ dependence as we calculated. In Ref [20], the $T^3$-dependence heat capacity is calculated to be 0.43fJ/K and 12fJ/K at temperature of 2K and 6K respectively. In the mentioned reference, the experimental values for these temperatures are 3.75fJ/K and 58fJ/K, which are 8.7 and 4.8 times larger than the calculated values. These ratios are approximately similar to our obtained total to bulk heat capacitance ratios. In addition, The temperature dependence of the experimental heat capacities in the considered range are in the form of $T^{2.49}$ and the deviation from the $T^3$ law can be attributed to the contribution of the surface free energy in the total heat capacity.

The variation of time constant with respect to the substrate thickness is presented in Fig. 4. This behavior is consistent with previously reported measurements on the time constant of NTC thermal sensors, which has analogous principle of operation and similar thermal model to that of transition-edge sensors [11].

In Fig. 5, the normalized frequency-dependent response of the bolometer versus the substrate thickness is depicted for the modulation frequencies of 1, 4, 10 and 50Hz. For this calculation, the total thermal conductance of the sample is taken to be approximately $10\text{-}50\mu WK^{-1}$ [21, 22]. The responsivity is normalized to its value at low thicknesses to show its trend when the thickness is changed.

It is noteworthy to mention that as the thickness of the substrate decreases, the responsivity is enhanced until it reaches a fairly constant value at the certain thickness (knee thickness) which is a frequency dependent parameter. Thus, thinning the substrate is desirable until it reaches the knee thickness, and below it, the responsivity would not improve as it is shown in Fig. 5. As it is illustrated in this figure, the knee thickness appears at higher substrate thicknesses for lower frequencies. It should also be noted that in addition to saturation of the responsivity at small values of the substrate thicknesses, it becomes constant for thicker substrates at frequencies near to zero (DC) modulation frequency. This is attributed to the slight effect of the heat capacitance at low frequencies, which is the only thickness dependent parameter in the calculation.

## IV. CONCLUSIONS

In summary, we presented a detailed investigation of the responsivity of the membrane-type bolometers when the substrate thickness is in the micrometer range. The failure of previous theories, which predict infinitive enhance of the responsivity to any desired value by thinning the substrate is taken cared by the proposed approach. In the presented work, it is shown that below the knee thickness, the surface effects could limit further increase of the responsivity. We modeled this limitation by the variation of the total specific heat with respect to the thickness of the substrate caused by the domination of the contribution of the surface energy in the total internal energy of the substrate. For thin substrates, the response shows a different behavior with respect to the thick substrate ones. By relating this anomaly to the surface effect and using the finite-size scaling method and calculating the surface energy, the bolometric response is calculated as a function of substrate thickness. The results for specific heat are consistent with previous empirical works on micrometer scale substrates devices.

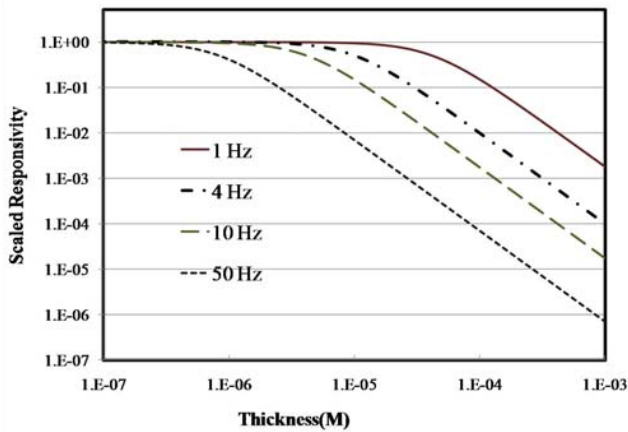

Fig. 5. Frequency response of transition edge superconductor versus substrate thickness for different modulation frequency.